\documentclass[twocolumn]{IEEEtran}
\usepackage{amsmath}
\usepackage{amsfonts}
\usepackage{cite}
\usepackage{bm}
\usepackage{url}
\usepackage{multibib}
\usepackage{xcolor} 
\usepackage{graphicx} 
\usepackage{tikz}
\usepackage{booktabs}

\DeclareMathOperator*{\argmin}{arg\,min}
\DeclareMathOperator*{\argmax}{arg\,max}

\begin{document}
\title{A Review of Convolutional Neural Networks for Inverse Problems in Imaging}

\author{Michael~T.~McCann,~\IEEEmembership{Member,~IEEE},
Kyong~Hwan~Jin,~
Michael~Unser,~\IEEEmembership{Fellow,~IEEE}
\thanks{Michael McCann is with the Center for Biomedical Imaging, Signal Processing Core and the Biomedical Imaging Group, EPFL, Lausanne, Switzerland (email: michael.mccann@epfl.ch).}
\thanks{K.H. Jin is with the Biomedical Imaging Group, EPFL, Lausanne, Switzerland.}
\thanks{K.H. Jin acknowledges the support from the ``EPFL Fellows'' fellowship program co-funded by Marie Curie from the European Union’s Horizon 2020 Framework Programme for Research and Innovation under grant agreement 665667.}
\thanks{Michael Unser is with the Biomedical Imaging Group, EPFL, Lausanne, Switzerland. 
}}

\pagebreak

\maketitle

\begin{abstract}
In this survey paper, we review recent uses of convolution neural networks (CNNs) to solve inverse problems in imaging.    
It has recently become feasible to train deep CNNs on large databases of images, and they have shown outstanding performance on object classification and segmentation tasks.
Motivated by these successes, researchers have begun to apply CNNs to the resolution of inverse problems such as denoising, deconvolution, super-resolution, and medical image reconstruction,
and they have started to report improvements over state-of-the-art methods, including sparsity-based techniques such as compressed sensing.
Here, we review the recent experimental work in these areas, with a focus on the critical design decisions: Where does the training data come from? What is the architecture of the CNN? and How is the learning problem formulated and solved?
We also bring together a few key theoretical papers that offer perspective on why CNNs are appropriate for inverse problems and point to some next steps in the field.
\end{abstract}

\section{Introduction}
\label{sec:intro}
The basic ideas underlying the use of convolutional neural networks (CNNs, also known as ConvNets) for inverse problems are not new.
Here, we give a very condensed history of CNNs to give context to what follows.
For more historical perspective, see \cite{schmidhuber_deep_2015}, and for an accessible introduction to deep neural networks and a summary of their recent history, see \cite{lecun_deep_2015}.
The CNN architecture was proposed in 1986 in \cite{rumelhart_learning_1986} and neural networks were developed for solving inverse imaging problems as early as 1988~\cite{zhou_image_1988}.
These approaches, which used networks with a few parameters and did not always include learning, were largely superseded by compressed sensing (or, broadly, convex optimization with regularization) approaches in the 2000s.
As computer hardware improved, it became feasible to train larger and larger neural networks, until, in 2012, Krizhevsky et al.~\cite{krizhevsky_imagenet_2012} achieved a significant improvement over the state of the art on the ImageNet classification challenge by using a GPU to train a CNN with 5 convolutional layers and 60 million parameters on a set of 1.3 million images. 
This work spurred a resurgence of interest in neural networks, and specifically CNNs, for not only computer vision tasks, but also inverse problems and more.

The purpose of this review is to summarize the recent works using CNNs for inverse problems in imaging; i.e., in problems most naturally formulated as recovering an image from a set of noisy measurements;
this criterion excludes detection, segmentation, classification, quality assessment, etc.
We also focus on CNNs, avoiding other architectures such as recurrent neural networks, fully-connected networks, and stacked denoising autoencoders.
We organized our literature search by application, looking for topics of broad interest where we could find at least three peer-reviewed papers from the last ten years.%
\footnote{
Much of the work on the theory and practice of CNNs is posted on the preprint server arXiv.org before eventually appearing in a traditional journal.
Because of the lack of peer review on arXiv.org, we have preferred not to cite these papers, except in cases where we are trying to illustrate a very recent trend or future direction for the field.
When citing a paper from arXiv, we follow the inline citation with an asterisk, e.g. [30]*.} 
The resulting applications and references are summarized in Table~\ref{tab:apps}.
The aim of this constrained scope is to allow us to draw meaningful generalizations from the surveyed works.

\begin{table}[htbp]
    \caption{Reviewed applications and associated references.}
    \centering
    \begin{tabular}{ccccc}
         denoising & deconvolution & super-resolution & MRI & CT  \\ \cmidrule(lr){1-1}\cmidrule(lr){2-2}\cmidrule(lr){3-3}\cmidrule(lr){4-4}\cmidrule{5-5}
         \cite{jain_natural_2009,burger_image_2012,xie_image_2012,wang_non-local_2016,chen_trainable_2016,zhang_beyond_2017} & 
         \cite{xu_deep_2014,schuler_learning_2016,chen_trainable_2016,schawinski_generative_2017} &
         \cite{cui_deep_2014,wang_deep_2015,kim_accurate_2016,kappeler_video_2016,wang_non-local_2016,dong_image_2016,li_video_2017} &
         \cite{wang_accelerating_2016,yang_deep_2016,oktay_multi-input_2016} &
         \cite{pelt_fast_2013,boublil_spatially-adaptive_2015,chen_low-dose_2017,jin2017deep}
    \end{tabular}

    \label{tab:apps}
\end{table}



The manuscript is organized as follows.
We begin in Section~\ref{sec:background} with a brief background on inverse imaging problems and how they can be formulated as learning problems.
We continue in Section~\ref{sec:results}, which summarizes the recent results obtained by using CNNs for a variety of image reconstruction applications.
We then survey the recent CNN-based reconstruction methods in detail in Section~\ref{sec:design}, with a focus on design decisions involving the training set, the network architecture, the formulation of the learning problem, and the optimization procedure itself.
We briefly cover some of the theoretical perspectives on the good performance of CNNs for inverse problems in Section~\ref{sec:theory}.
We then discuss critiques of the CNN-based approach in Section~\ref{sec:critiques} and conclude in Section~\ref{sec:future} with our view of the future directions for the field.

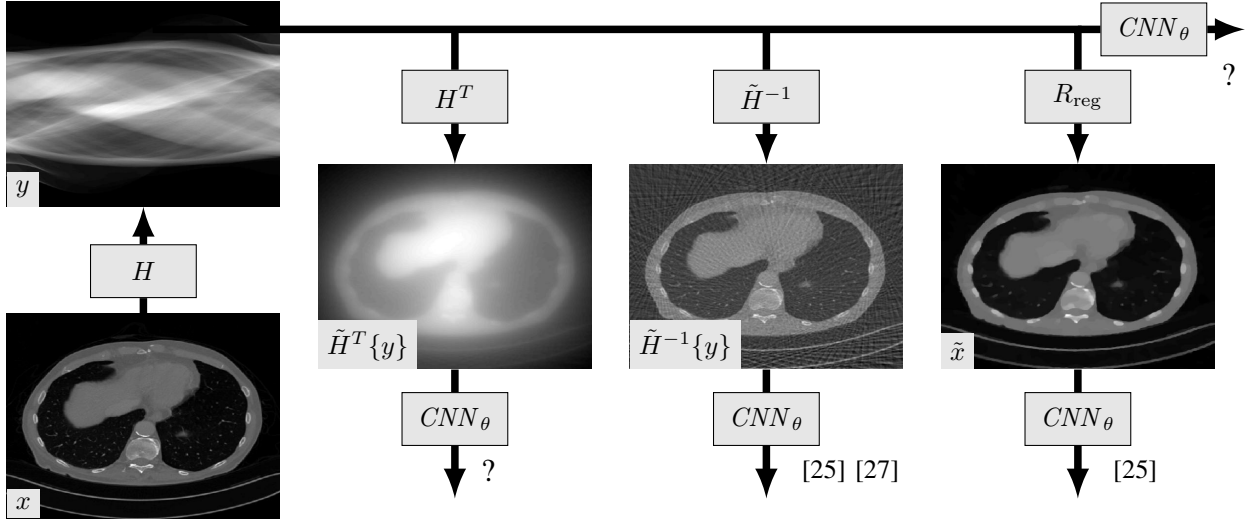
\begin{figure*}[htbp]
\centering

\newlength{\imWidth}
\setlength{\imWidth}{.2\textwidth}

\tikzset{>=latex}
\begin{tikzpicture}
\usetikzlibrary{positioning,calc}
\tikzstyle{block} = [draw, fill=gray!20, rectangle, 
    minimum height=2em, minimum width=4em];
\tikzstyle{image} = [inner sep=0pt];
\tikzstyle{label} = [fill=gray!20, rectangle,
minimum height = 1em, minimum width = 1em, anchor = south west];

\node [image] (input) {\includegraphics[width=\imWidth]{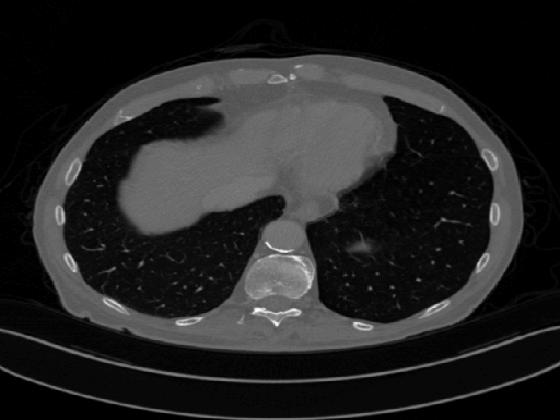}};
\node [block, above = .2 of input] (H) {$H$};
\node [image, above = .5 of H] (sino) {\includegraphics[width=\imWidth]{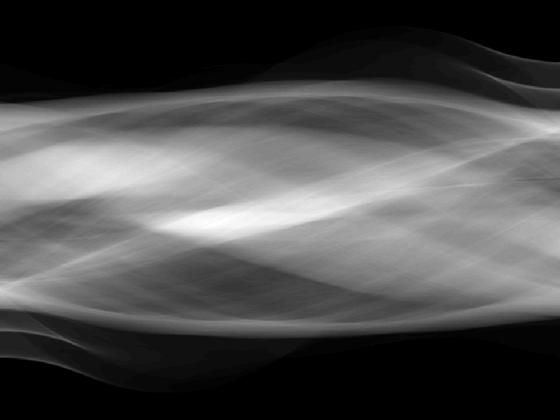}};

\node [image, above right = -.75 and .5cm of input] (HTg) {\includegraphics[width=\imWidth]{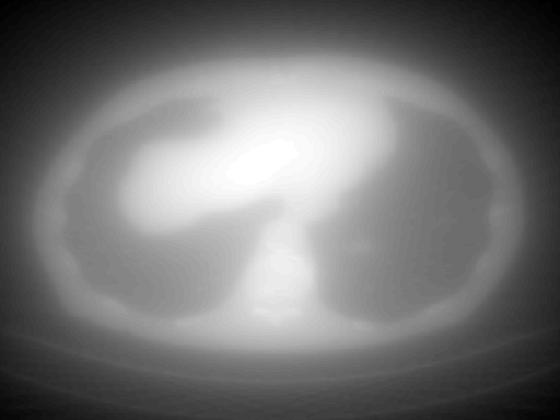}};

\node [image, right = .5cm of HTg] (FBP) {\includegraphics[width=\imWidth]{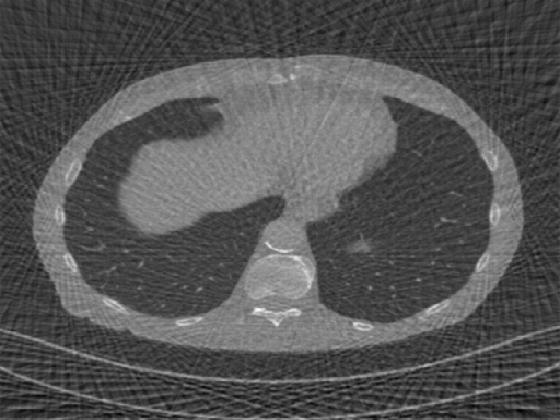}};

\node [image, right = .5cm of FBP] (TV) {\includegraphics[width=\imWidth]{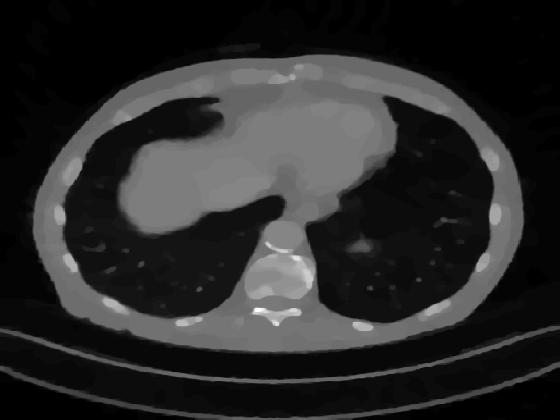}};

\node [above = -.5 of sino]  (sinoTop) {};
\node [coordinate] (join1) at (sinoTop -| HTg) {};
\node [coordinate] (join2) at (sinoTop -| FBP) {};
\node [coordinate] (join3) at (sinoTop -| TV) {};

\node [block] (HT) at ($(join1)!0.5!(HTg.north)$) {$H^T$};
\node [block] (Hinv) at ($(join2)!0.5!(FBP.north)$) {$\tilde{H}^{-1}$};
\node [block] (R) at ($(join3)!0.5!(TV.north)$) {$R_{\rm reg}$};

\node [block, below = .3cm of HTg] (CNN1)  {$\mathit{CNN}_\theta$};
\node [block, below = .3cm of FBP] (CNN2) {$\mathit{CNN}_\theta$};
\node [block, below = .3cm of TV] (CNN3) {$\mathit{CNN}_\theta$};
\node [block, right = .3cm of join3] (CNN4) {$\mathit{CNN}_\theta$};

\node [below = .7cm of CNN1] (q1) {};
\node [above right = .1cm and .1cm of q1] {\large ?};
\node [below = .7cm of CNN2] (q2) {};
\node [above right = .1cm and .1cm of q2] {\cite{boublil_spatially-adaptive_2015}\cite{jin2017deep}};
\node [below = .7cm of CNN3] (q3) {};
\node [above right = .1cm and .1cm of q3] {\cite{boublil_spatially-adaptive_2015}};
\node [right = .5cm of CNN4] (q4) {};
\node [below left = .2cm and 0cm of q4] {\large ?};

\node [label] at (input.south west) {$x$};
\node [label] at (sino.south west) {$y$};
\node [label] at (HTg.south west) {$\tilde{H}^T \{y\}$};
\node [label] at (FBP.south west) {$\tilde{H}^{-1} \{y\}$};
\node [label] at (TV.south west) {$\tilde{x}$};

\draw[->,line width=1mm] (input) -- (H) -- (sino);
\draw[->,line width=1mm] (join1) -- (HT) -- (HTg);
\draw[->,line width=1mm] (join2) -- (Hinv) --(FBP);
\draw[->,line width=1mm] (sinoTop) -- (join3) -- (R) -- (TV);

\draw[->,line width=1mm] (HTg) -- (CNN1) -- (q1);
\draw[->,line width=1mm] (FBP) -- (CNN2) -- (q2);
\draw[->,line width=1mm] (TV) -- (CNN3) -- (q3);
\draw[->,line width=1mm] (join3) -- (CNN4) -- (q4);

\end{tikzpicture}
\caption{Block diagram of image reconstruction methods, using images from X-ray CT as examples.
An image, $x$ creates measurements, $y$, that can be used to estimate $x$ in a variety of ways.
The traditional approach is to apply a direct inversion, $\tilde{H}^{-1}$, which is artifact-prone in the sparse-measurement case (note the stripes in the reconstruction).
The current state of the art is a regularized reconstruction, $R_{\rm reg}$, written in general in \eqref{eq:obj_reg}.
Several recent works apply CNNs to the result of the direct inversion or an iterative reconstruction, but it might also be reasonable to use as input the measurements themselves or the back projected measurements.
}
\label{fig:CT_example}
\end{figure*}

\section{Background}
\label{sec:background}
We begin by introducing inverse problems and contrasting the traditional approach to solving them with a learning-based approach.
For a textbook treatment of inverse problems, see~\cite{kirsch2011introduction}.
Throughout the section, we use X-ray CT as a running example, and Figure~\ref{fig:CT_example} shows images of the various mathematical quantities we mention.

\subsection{Learning for Inverse Problems in Imaging}
\label{sec:inv_and_learning}

Mathematically speaking, an imaging system is an operator $H:\mathcal{X} \rightarrow \mathcal{Y}$ that acts on an image $x \in \mathcal{X}$, to create a vector of measurements $y \in \mathcal{Y}$, with $H\{x\} = y$.
The underlying function/vector spaces are
\begin{itemize}
\item the space, $\mathcal{X}$, of {\bf acceptable images} which can be 2D, 3D, or even 3D+time, with its values  representing a physical quantity of interest, such as X-ray attenuation or concentration of fluorophores; and
\item the space, $\mathcal{Y}$, of {\bf measurement vectors} which depends on the imaging operator and could include images (discrete arrays of pixels), Fourier samples, line integrals, etc.
\end{itemize}
We typically consider $x$ to be a continuous object (function of space), while $y$ is usually discrete: $\mathcal{Y} = \mathbb{R}^M$. 
For example, in X-ray CT, $x$ is an image representing X-ray attenuations, $H$ represents the physics of the X-ray source and detector, and $y$ is the measured sinogram (see Figure~\ref{fig:CT_example}).

In an inverse imaging problem, we aim to develop a reconstruction algorithm (which is also an operator), $R:\mathcal{Y} \rightarrow \mathcal{X}$ in order to recover the original image, $x$, from the measurements, $y$.
The dominant approach for reconstruction, which we call the \emph{objective function approach}, is to model $H$ and recover an estimate of $x$ from $y$ by
\begin{equation}
\label{eq:obj}
R_{\rm obj}\{y\} = \argmin_{x \in \mathcal{X}} f(H\{x\}, y),
\end{equation}
where $H:\mathcal{X} \rightarrow \mathcal{Y}$ is the system model, which is usually linear, and $f:\mathcal{Y}\times\mathcal{Y} \rightarrow \mathbb{R}^+$ is an appropriate measure of error.
Continuing the CT example, $H$ would be a discretization of the X-ray transform (such as Matlab's \texttt{radon}) and $f$ could be the Euclidean distance, $\|H\{x\}-y\|_2$.
For many applications, decades of engineering have gone into developing a fast and reasonably accurate inverse operator, $\tilde{H}^{-1}$, so Eq. \eqref{eq:obj} is easily solved with $R_{\rm obj}\{y\} = \tilde{H}^{-1}\{y\}$;
for CT, $\tilde{H}^{-1}$ is the filtered back projection (FBP) algorithm.
An important, related operator is the back projection, $H^T:\mathcal{Y} \rightarrow \mathcal{X}$, which can be interpreted as the simplest way to put measurements back into the image domain (see Figure~\ref{fig:CT_example}).

These direct inverses begin to show significant artifacts when the number or quality of the measurements decreases, either because the underlying discretization breaks down, or because the inversion of \eqref{eq:obj} becomes ill-posed (lacking a solution, lacking a unique solution, or being unstable with respect to the measurements).
Unfortunately, in many real-world problems, measurements are costly (in terms of time, or, e.g., X-ray damage to the patient), which motivates us to collect as few as possible.
In order to reconstruct from sparse or noisy measurements, it is often better to use a regularized formulation,
\begin{equation}
\label{eq:obj_reg}
R_{\rm reg}\{y\} = \argmin_{x \in \mathcal{X}} f \left(H\{x\}, y \right) + g(x),
\end{equation}
where $g :\mathcal{X} \rightarrow \mathbb{R}^+ $ is a regularization functional that promotes solutions that match our prior knowledge of $x$, and, simultaneously, makes the problem well-posed.
For CT, $g$ could be the total variation (TV) regularization, which penalizes large gradients in $x$.

From this perspective, the challenge of solving an inverse problem is designing and implementing \eqref{eq:obj_reg} for a specific application.
Much effort has gone into designing general-purpose regularizers and minimization algorithms.
For example, compressed sensing~\cite{candes_robust_2006} provides sparsity-promoting regularizers.
Nonetheless, in the worst case, a new application necessitates developing accurate and efficient $H$, $g$, and $f$, along with a minimization algorithm.


An alternative to the objective function approach is the \emph{learning approach}, where a training set of ground truth images and their corresponding measurements, $\{(x_n, y_n)\}_{n=1}^{N}$, is known.
A parametric reconstruction algorithm, $R_{\rm learn}$, is then learned by solving
\begin{equation}
R_{\rm learn} = \argmin_{R_\theta, \theta \in \Theta} \sum^N_{n=1} f(x_n, R_\theta\{y_n\}) + g(\theta),
\label{eq:learn_reg}
\end{equation}
where $\Theta$ is the set of all possible parameters,
$f:\mathcal{X}\times\mathcal{X} \rightarrow \mathbb{R}^+$ is a measure of error, and $g:\Theta \rightarrow \mathbb{R}^+$ is a regularizer on the parameters with the aim of avoiding overfitting.
Once the learning step is complete, $R_{\rm learn}$ can then be used to reconstruct a new image from its measurements.

To summarize, in the objective function approach, the reconstruction function is itself a regularized minimization problem, while in the learning approach, the solution of a regularized minimization problem is a parametric function that can be used to solve the inverse problem.
The learning formulation is attractive because it overcomes many of the limitations of the objective function approach: there is no need to handcraft the forward model, cost function, regularizer, and optimizer from \eqref{eq:obj_reg}.
On the other hand, the learning approach requires a training set, and the minimization \eqref{eq:learn_reg} is typically more difficult than \eqref{eq:obj_reg} and requires a problem-dependant choice of $f$, $g$, and the class of functions described by $R$ and $\Theta$.

Finally, we note that the learning and objective function approaches describe a spectrum rather than a dichotomy.
In fact, the learning formulation is strictly more general, including the objective function formulation as a special case.
As we will discuss further in Section~\ref{sec:arch}, which (if any) aspects of the objective formulation approach to retain is a critical choice in the design of learning-based approaches to inverse problems in imaging.

\begin{figure*}[htbp]
\centering
\includegraphics[width=13cm,trim={0cm 1.9cm 0cm 2.2cm},clip]{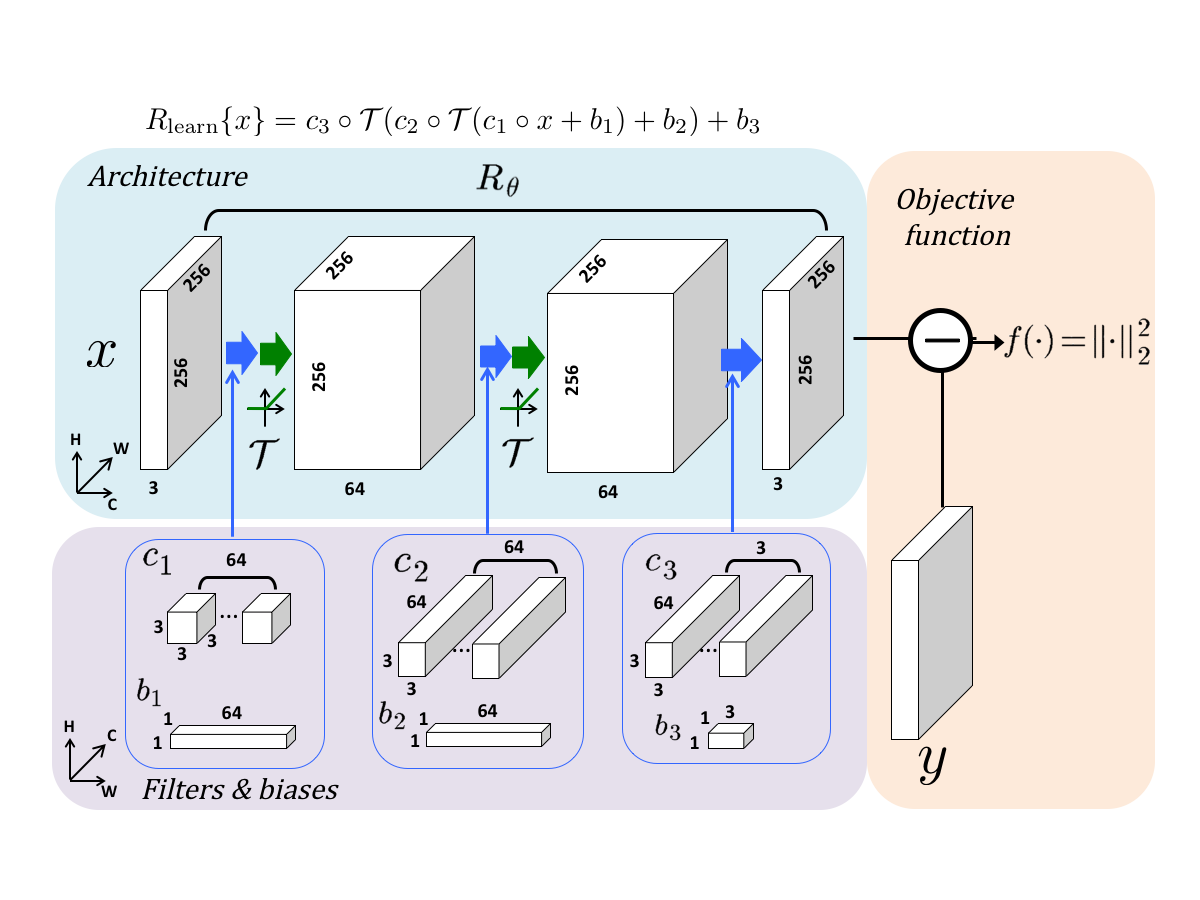}
\caption{Illustration of a typical CNN architecture for $256^2$ px RGB images, including the objective function used for training.
$\mathcal{T}(\cdot)$ is the ReLU function (point-wise nonlinear function). $\circ$ denotes a 2-D convolution.
The convolutions in each layer is described by a 4-D tensor representing a stack of 3D filters.}
\label{fig:cnn_str}
\end{figure*}

\subsection{Convolutional Neural Networks}
Our focus here is the formulation of \eqref{eq:learn_reg} using CNNs.
Using a CNN means, roughly, fixing the set of functions, $R_\theta$, to be a sequence of filtering operations alternating with simple nonlinear operations.
This class of functions is parametrized by the values of the filters used (also known as \emph{filter weights}), and these filter weights are the parameters over which the minimization occurs.
For illustration, Figure~\ref{fig:cnn_str} shows a typical CNN architecture.

We will describe some of the theoretical motivations for using CNNs as the learning architecture for inverse problems in Section~\ref{sec:theory}, but we mention some practical advantages here.
First, the forward operation of a CNN consists of (usually small) convolutions and simple, pointwise nonlinear functions.
This means that once training is complete, the execution of $R_{\rm learn}$ is very fast and amenable to hardware acceleration on GPUs.
Second, the gradient of \eqref{eq:learn_reg} is computable via the chain rule, and these gradients again involve small convolutions, meaning that the parameters can be learned efficiently via gradient descent.

When the first CNN-based method entered the ImageNet Large Scale Visual Recognition Challenge in 2012~\cite{krizhevsky_imagenet_2012},
its error rate on the object localization and classification task was 15.3\%, as compared to an error rate 26.2\% for the next closest method and 25.8\% for the winner from 2011.
In subsequent competitions (2013-2016), the majority of the entries (and all of the winners) were CNN-based and continued to improve substantially, with the 2016 winner achieving an error rate of just 2.99\%.
Can we expect such large gains in inverse problems?
That is, can we expect denoising results to improve by an order of magnitude (20 dB) in the next few years?
In the next section, we answer this question by surveying the results reported by recent CNN-based approaches to image reconstruction.
 

\section{Current State of Performance}
\label{sec:results}

Of the inverse problems we review here, denoising provides the best look at recent trends in results because there are standard experiments that appear in most papers.
Work on CNN-based denoising from 2009~\cite{jain_natural_2009} showed an average PSNR of 28.5 on the Berkeley Segmentation Dataset, a less than 1 dB improvement over contemporary wavelet and Markov random field-based approaches.
For comparison, one very recent denoising work~\cite{zhang_beyond_2017} reported a 0.7 dB improvement on a similar experiment, which remains a less than 1 dB better than contemporary non-CNN methods (including BM3D, which had remained the state-of-the-art for years).
As another point of reference, in 2012, one CNN approach~\cite{burger_image_2012} reported an average PSNR of 30.2 dB on a set of standard test images (Lena, peppers, etc.), less than 0.1 dB better than comparisons, and another~\cite{xie_image_2012}, reported an average of 30.5 dB on the same experiment.
The recent \cite{zhang_beyond_2017} achieves an average of 30.4 dB under the same conditions.
One important perspective on these denoising results is that the CNN is learning the distribution of natural images (or equivalently, is learning a regularization).
Such a CNN could be reused inside an iterative optimization as a proximal operator to enforce this learned regularization for any inverse problem.

The trends are similar in deblurring and super-resolution, though experiments are more varied and therefore harder to compare.
For deblurring, \cite{xu_deep_2014} showed around a 1 dB PSNR improvement over comparison methods, and \cite{schuler_learning_2016} showed a further improvement of around 1 dB.
For super-resolution, work from 2014 \cite{cui_deep_2014} reported a less than 0.5 dB improvement in PSNR over comparisons.
In the next two years, \cite{wang_deep_2015} and \cite{dong_image_2016} both reported a 0.5 dB PSNR increase over this baseline.
Even more recent work, \cite{ledig_photo-realistic_2016}*, improves the 2014 work by around 1.5 dB in PSNR.
For video super-resolution, \cite{kappeler_video_2016} improves on non-CNN-based methods by about 0.5 dB PSNR and \cite{li_video_2017} improves upon that result by another 0.5 dB.

\begin{figure*}[htbp]
\centering
\includegraphics[trim = 0mm 50mm 0mm 40mm,clip=true,width=16.5 cm]{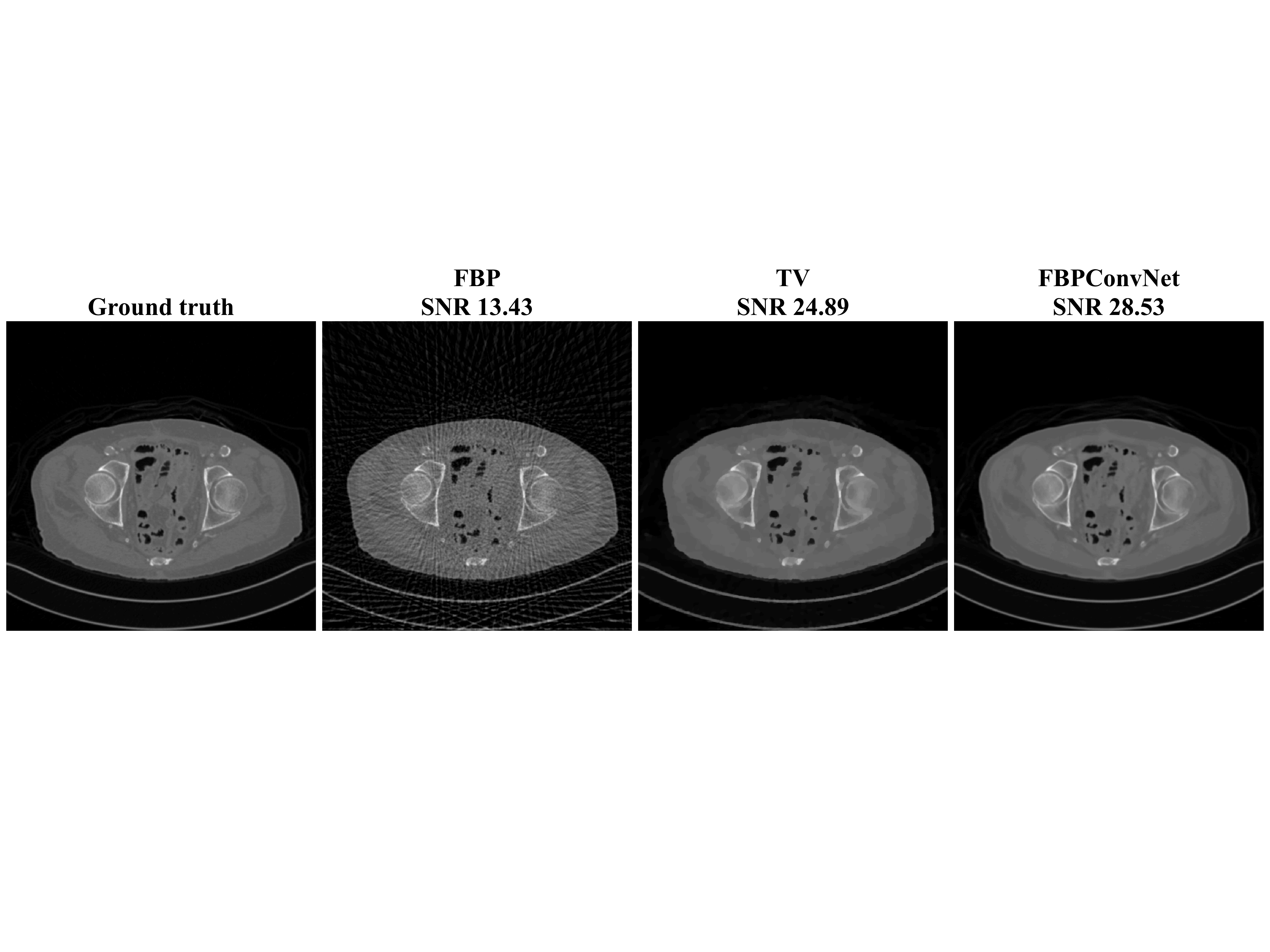}
\includegraphics[trim = 0mm 50mm 0mm 49.5mm,clip=true,width=16.5 cm]{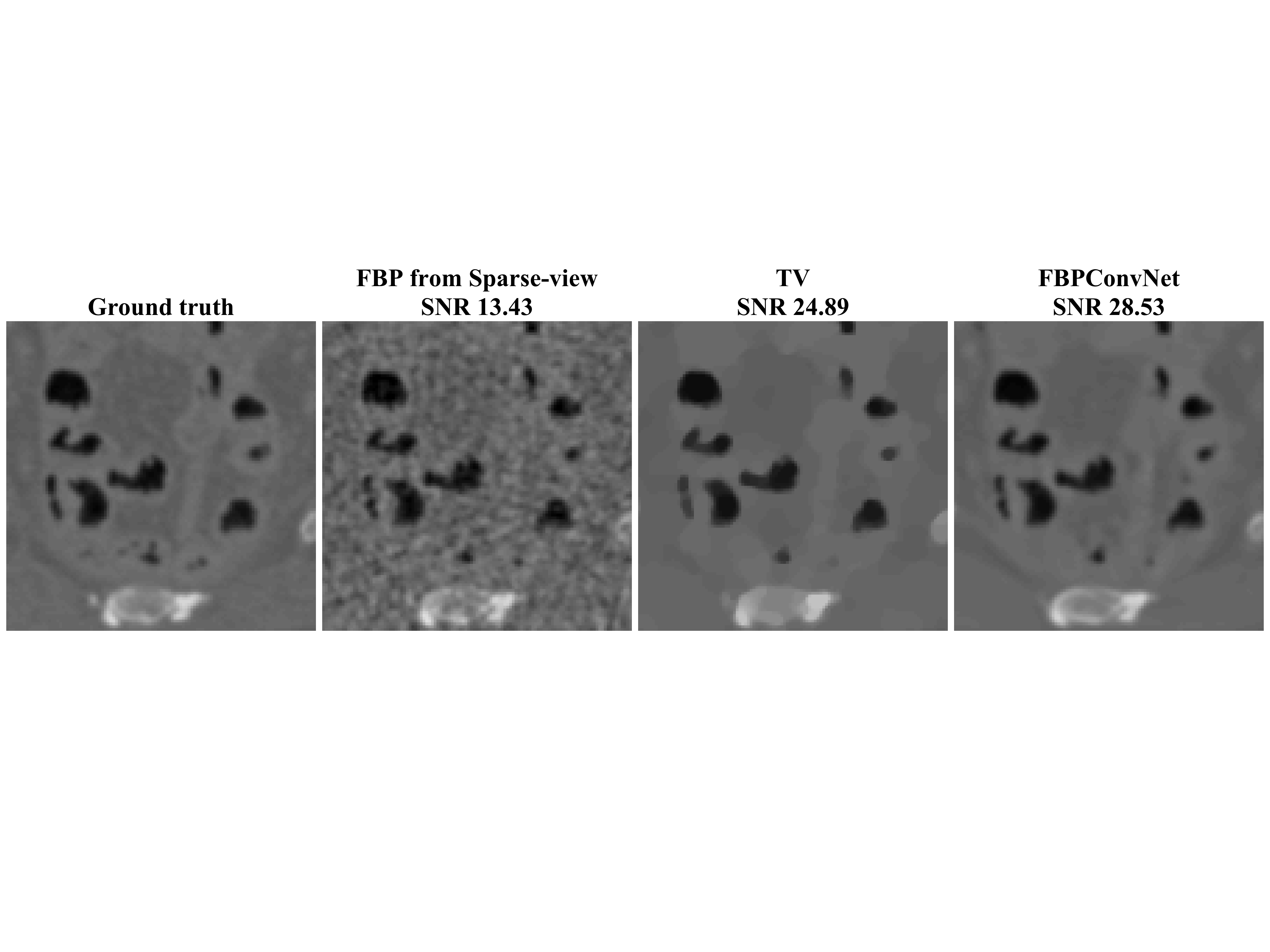}
\caption{An example of X-ray CT reconstructions.
The ground truth (left column) comes from an FBP reconstruction using 1000 views.
The next three columns show reconstructions from just 50 views using FBP, a regularized reconstruction, and from a CNN-based approach (images reproduced with permission from \cite{jin2017deep}). 
The CNN-based reconstruction preserves more of the texture present in the ground truth and results in a significant increase in SNR.
}
\label{fig:res_biomed}
\end{figure*}

For inverse problems in medical imaging, direct comparison between works is impossible due to the wide variety of experimental setups.
A 2013 CNN-based work~\cite{pelt_fast_2013} shows improvement in limited-view CT reconstruction over direct methods and unregularized iterative methods, but does not compare to regularized iterative methods.
In 2015, \cite{boublil_spatially-adaptive_2015} showed in full-view CT an improvement of several dB in SNR over direct reconstruction and around 1 dB improvement over regularized iterative reconstruction.
Recently, \cite{chen_low-dose_2017} showed about 0.5 dB improvement in PSNR over TV-regularized reconstruction, while \cite{jin2017deep} showed a larger (1-4 dB) improvement in SNR over a different TV-regularized method (Figure~\ref{fig:res_biomed}).
In MRI, \cite{yang_deep_2016} demonstrates performance equal to the state-of-the-art, with advantages in running time.

Do these improvements matter?
CNN-based methods have not, so far, had the profound impact on inverse problems that they have for object classification.
Indeed, the difference between 30 and 30.5 dB is impossible to see by eye.
On the other hand, these improvements occur in heavily studied fields: we have been denoising the Lena image since the 1970s.
Further, CNNs offer some unique advantages over many traditional methods.
The design of the CNN architecture can be more or less decoupled from the application at hand and can be reused from problem to problem.
They can also be expanded in straightforward ways as computer memory grows and there is some evidence that larger networks lead to better performance.
Finally, once trained, running the model is fast (dozens of convolutions per image, usually less than one second).
This means that CNN-based methods can be attractive in terms of running time even if they do not improve upon state-of-the-art performance.

\begin{figure*}[htb]
\centering
\includegraphics[trim = 0mm 80mm 0mm 80mm,clip=true,width=12cm]{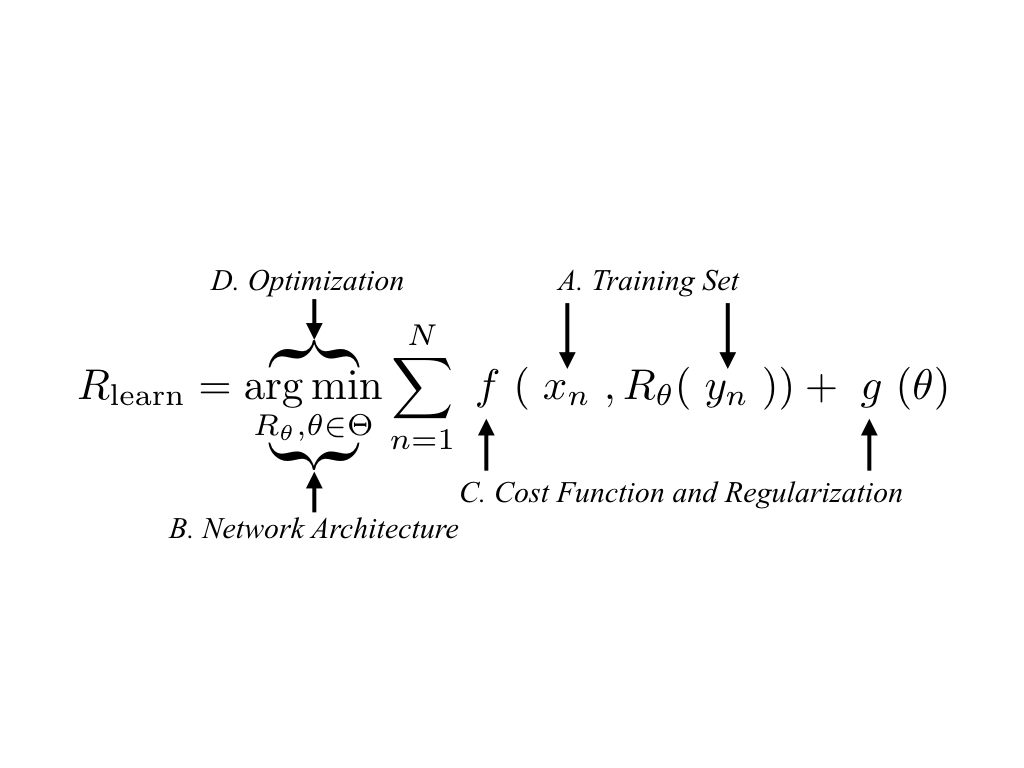}
\caption{The learning equation, repeated from the introduction, which we use to organize the parts of Section~\ref{sec:design}.}
\label{fig:eq_smmary}
\end{figure*}

\section{Designing CNNs for Inverse Problems}
\label{sec:design}
In this section, we survey the design decisions needed to develop CNN-based approaches for inverse problems in imaging.
We organize the section around the learning equation as summarized in Figure~\ref{fig:eq_smmary}, first describing how the training set is created, then how the network architecture is designed, and then how the learning problem is formulated and solved.

\subsection{Training Set}
\label{sec:training}

Learning requires a suitable training set, i.e. the (input, output) pairs from which the CNN will learn.
In a typical learning problem, training outputs are provided by some oracle labeling a set of inputs.
For example, in object classification, a set of human graders might view a large number of images and provide annotations for each.
In the inverse problem setting, this is considerably more difficult because no such oracle exists.
For example, in X-ray CT, to generate a training set we would need to image a large number of physical phantoms for which we have exact 3D models, which is not feasible in practice.
The choice of the training set also constrains the network architecture because the input and output of the network must match the dimensions of $y_n$ and $x_n$, respectively.

\subsubsection{Generating Training Data}
In some cases, generating training data is straightforward because the forward model we aim to invert is known exactly and easily computable.
In denoising, training data is generated by corrupting images with noise;
the noisy image then serves as training input and the clean image as the training output, as in, e.g., \cite{jain_natural_2009,burger_image_2012}.
Or, the noise itself can serve as the oracle output, in a scheme called residual learning~\cite{oktay_multi-input_2016}, \cite{zhang_beyond_2017}.
super-resolution follows the same pattern, where training pairs are easily generated by downsampling, as in, e.g., \cite{dong_image_2016}.
The same is true for deblurring, where training pairs can be generated by blurring~\cite{xu_deep_2014,schawinski_generative_2017,schuler_learning_2016}.

In medical imaging, the focus is on reconstructing from real measurements and the corresponding ground truth is not usually known.
The emerging paradigm is to learn to reconstruct from sparse measurements, using reconstructions from fully-sampled measurements to train. 
For example, in MRI reconstruction, \cite{yang_deep_2016} trains using under-sampled k-space data as inputs and reconstructions from fully-sampled k-space data as outputs.
Likewise, \cite{jin2017deep} uses a low-view CT reconstruction as input and a high-view CT reconstruction as output.
Or, the CNN can learn from low-dose (noisy) measurements~\cite{boublil_spatially-adaptive_2015}.

\subsubsection{Preprocessing}\label{sec:preprocessing}
Another aspect of training data preparation is whether the training inputs are the measurements themselves, or whether some preprocessing occurs.
In denoising, it is natural to use the raw measurements, which are of the same dimensions as the reconstruction.
But, in the other applications, the trend is to use a direct inverse operator to preprocess the network input.
Following the notation in Section~\ref{sec:inv_and_learning}, this can be viewed as a combination of the objective function and learning approach, where instead of $R_{\rm learn}$ being a CNN, it is the composition of a CNN with a direct inverse: $R_\theta \circ \tilde{H}^{-1}$.
For example, 
in super-resolution, \cite{wang_deep_2015,dong_image_2016,kappeler_video_2016} first upsample and interpolate the low-resolution input images; in CT, \cite{jin2017deep} and 
\cite{boublil_spatially-adaptive_2015} preprocess with the FBP, \cite{boublil_spatially-adaptive_2015} also preprocesses with an iterative reconstruction; and, in MRI, ~\cite{wang_accelerating_2016} preprocesses with the inverse Fourier transform.

Without preprocessing, the CNN must learn the underlying physics of the inverse problem.
It is not even clear that this is possible with CNNs (e.g., what is the meaning of filtering an X-ray CT sinogram?).
Preprocessing is also a way to leverage the significant engineering effort that has gone into designing these direct inverses over the past decades.
Superficially, this type of preprocessing appears to be inversion followed by denoising, which is a standard, if ad hoc, approach to inverse problems.
What is unique here is that the artifacts caused by direct inversion, especially in the sparse measurement case, are usually highly structured and therefore not good candidates for generic denoising approaches.
Instead, the CNN is allowed to learn the specific character of these artifacts.

A practical aspect of preprocessing is controlling the dynamic range of the input.
While not typically a problem when working with natural images or standardized datasets, there may be huge fluctuations in the intensity or contrast of the measurements in certain inverse problems.
To avoid a small set of images dominating the error during training, it is best to scale the dynamic range of the training set~\cite{jin2017deep,oktay_multi-input_2016}.
Similarly, it may be advantageous to discard training patches without sufficient contrast.

\subsubsection{Training Size}
CNNs typically have at least thousands of parameters to train;
thus, the number of (input, output) pairs in the training set is of important practical concern.
The number of training pairs varied among the papers we surveyed.
The biomedical imaging papers tended to have the fewest samples (e.g., 500 brain images~\cite{wang_accelerating_2016} or 2000 CT images~\cite{pelt_fast_2013}),
while papers on natural images had the most (e.g., pretraining on 395,909 natural images~\cite{li_video_2017}).

A further complication is that, depending on the network architecture, images may be split into patches for training.
Thus, depending on the dimensions of the training images and the chosen patch size, numerous patches can be created from a small training set.
The patch size also has important ramifications for the performance of the network and is linked to its architecture, with larger filters and deeper networks requiring larger training patches~\cite{kim_accurate_2016}.

With a large CNN and a small training set, overfitting must be avoided by regularization during learning and/or the use of a validation set (e.g., \cite{pelt_fast_2013}) (discussed more in Sections~\ref{sec:loss} and \ref{sec:optimization}).
These strategies are necessary to produce a CNN that generalizes at all, but they do not overcome the fact that the performance of the CNN will be limited by the size and variety of the training set.
One strategy to increase the training set size is data augmentation, where new (input, output) pairs are generated by transforming existing ones.
For example, \cite{li_video_2017} augmented training pairs by scaling them in space and time, turning 20,000 pairs into 70,000 pairs.
The augmentation must be application-specific because the trained network will be approximately invariant to the transforms used.
Another strategy to effectively increase the training set size is to use a pretrained network.
For example, \cite{kappeler_video_2016} first trains a CNN for image super-resolution with a large image dataset, then retrains with videos.




\subsection{Network Architecture}
By network architecture, we mean the choice of the family of CNNs, $R_\theta$ parameterized by $\theta$.
In our notation, $R_\theta$ represents a CNN with a specific architecture while $\theta$ are the weights to be learned during the training.
\label{sec:arch}
There is great variety among CNN-based methods regarding their architecture: how many convolutional layers, what filter sizes, which nonlinearities, etc.
For example, \cite{dong_image_2016} uses 8,032 parameters, while \cite{li_video_2017} uses on the order of one hundred thousand. 
In this section, we survey recent approaches to CNN architecture design for inverse problems.

The simplest approach to architecture design is simply stack of series of convolutional layers and non-linear functions~\cite{chen_low-dose_2017,chen_trainable_2016}, see Figure~\ref{fig:cnn_str}.
This provides a baseline to check the feasibility of the network for the given application.
It is straightforward to adjust the size of such a network, either by changing the number of layers, the number of channels per layer, or the size of the filters in each layer.
For example, keeping the filters small ($3 \times 3 \times 3$ px) allows the network to be deeper for a given number of parameters~\cite{oktay_multi-input_2016};
constraining the filters to be separable~\cite{xu_deep_2014} further reduces the number of parameters.
Doing this can give the experimenter a sense of the training time required on their hardware as well as the effects of the network size on performance.
From this simple starting point, the architecture can be tweaked for greater performance; for example, by adding downsampling and upsampling operations~\cite{jin2017deep}, 
or by simply adding more layers~\cite{li_video_2017}.


Instead of using ad hoc architecture design, one can adapt a successful CNN architecture from another application.
For example, \cite{jin2017deep} adapts a network designed for biomedical image segmentation to CT reconstruction by changing the number of output layers from two (background and foreground images) to one (reconstructed image).
These architectures can also be connected end-to-end, creating modular or hierarchical designs.
For example, a four-times super-resolution architecture can be created by connecting two two-times super-resolution networks~\cite{wang_deep_2015}.
This is distinct from training a two-times super-resolution network and applying it twice because the two modules of the CNN are trained as a unit.

A second approach is to begin with an iterative optimization algorithm and unroll it, turning each iteration into a layer of a network.
In such a scheme, filters that are normally fixed in the iterative minimization are instead learned.
The approach was pioneered in \cite{gregor_learning_2010}, for sparse coding;
their results showed that the learned algorithms could achieve a given error in fewer iterations than the standard ones.
Because many iterative optimization algorithms alternate filtering steps (linear updates) with pointwise nonlinear steps (proximal/shrinkage operations), the resulting network is often a CNN.
This was the approach in \cite{yang_deep_2016}, where the authors unrolled the ADMM algorithm to design a CNN for MRI reconstruction, with state-of-the-art results and improvements in running time.
For networks designed in this way, the original algorithm is a specific case and, therefore, the performance of the network cannot be worse than the original algorithm if training is successful.
The concept of unrolling can also be applied at a coarser scale, as in \cite{schuler_learning_2016}, where the modules of the network mimic the steps of a typical blind deconvolution pipeline: extract features, estimate kernel, estimate image, repeat.

Another promising design approach, similar to unrolling, is to learn only some part of an existing iterative method.
For example, given the modular nature of popular iterative optimization schemes such as the ADMM, a CNN can be employed as a proximal (denoising) operator while the rest of the algorithm remains unchanged~\cite{zhang_learning_2017}*.
This design combines many of the good aspects of both the objective function and learning-based approaches, and allows a single CNN to be used for several different inverse problems without retraining. 


\subsection{Cost Function and Regularization}
\label{sec:loss}

In this section, we survey the approaches taken to actually train the CNN, including the choice of a cost function, $f$, and regularizer, $g$.
For a textbook coverage of the subject of learning, see \cite{mitchell1997machine}.

Understanding the learning minimization problem as a statistical inference can provide useful insight into the selection of the cost and regularization functions.
From this perspective, we can formulate the goal of learning as maximizing the conditional likelihood of each training output given the corresponding training input and CNN parameters,

\begin{equation*}
\begin{split}
&\text{given} \quad \{(x_n, y_n)\}_{n=1}^{N}, \\
& R_{\rm learn} = \argmax_{{R_\theta, \theta \in \Theta}} \prod^N_{n=1} P(y_n \mid x_n, \theta),
\end{split}
\end{equation*}
where $P$ is a conditional likelihood.
When this likelihood follows a Gaussian distribution, this optimization is equivalent to the one from the introduction, \eqref{eq:learn_reg}, with $f$ being the Euclidean distance and no regularization.
Put another way, learning with the standard, Euclidean cost and no regularization implicitly assumes a Gaussian noise model;
this is a well-known fact in inverse problems in general. 
This formulation is used in most of the works we surveyed, \cite{boublil_spatially-adaptive_2015,dong_image_2016,jain_natural_2009,xu_deep_2014,oktay_multi-input_2016,kappeler_video_2016,zhang_beyond_2017,chen_low-dose_2017,burger_image_2012}, despite the fact that several raise questions about whether it is the best choice~\cite{boublil_spatially-adaptive_2015,zhao_loss_2017}.

An alternative is the \emph{maximum a posteriori} formulation, which maximizes the joint probability of the training data and the CNN parameters, which can be decomposed into several terms using Bayes rule,
\begin{equation}
\begin{split}
&\text{given} \quad \{(x_n, y_n)\}_{n=1}^{N}, \\
& R_{\rm learn} = \argmax_{{R_\theta, \theta \in \Theta}} \prod^N_{n=1} P(y_n \mid x_n, \theta)  P(\theta).
\end{split}
\end{equation}
This formulation explicitly allows prior information about the desired CNN parameters, $\theta$, to be used.
Under a Gaussian model for the weights of the CNN as well as the noise, this formulation results in a Euclidean cost function and a Euclidean regularization on the weights of the CNN, $g(\theta) = \sigma^{-2} \|\theta\|_2^2$.
Other examples of regularizations for CNNs are the total generalized variation norm or sparsity on the coefficients.
Regularized approaches are taken in \cite{yang_deep_2016,wang_accelerating_2016,cui_deep_2014,chen_trainable_2016}.

\subsection{Optimization}
\label{sec:optimization}
Once an objective function for learning has been fixed, it remains to actually minimize it.
This is a crucial and deep topic, but, from the practical perspective, it can be treated as a black box due to the availability of several high-quality software libraries that can perform efficient training of user-defined CNN architectures.
For a comparison of these libraries, refer to \cite{bahrampour_comparative_2015}*; here, we provide a basic overview.

The popular approaches to CNN learning are variations on gradient descent.
The most common is stochastic gradient descent (SGD), used in ~\cite{wang_deep_2015,boublil_spatially-adaptive_2015}, where, at each iteration, the gradient of the cost function is computed using random subsets of the available training.
This reduces the overall computation compared to computing the true gradient, while still providing a good approximation.
The process can be further tuned by adding momentum, i.e., combining gradients from previous iterations in clever ways, or by using higher order gradient information as in BFGS~\cite{yang_deep_2016}.

Initial weights can be set to zero, or chosen from some random distribution (Gaussian or uniform).
Because learning is nonconvex, the initialization does potentially change which minimum the network converges to, but, not much difference is observed in practice.
However, good initializations can improve the speed of convergence.
This explains the popularity of taking pretrained networks, or, in the case of an unrolled architecture, initializing the network weights based on corresponding known filters. 
Recently, a procedure called \emph{batch normalization}, where the inputs to each layer of the network are normalized, was proposed as a way to increase learning speed and reduce sensitivity to initialization~\cite{ioffe2015batch}.

As mentioned is Section~\ref{sec:training}, overfitting is a serious risk when training networks with potentially millions of parameters.
In addition to augmenting the training set, steps can be taking during training to reduce overfitting.
The simplest is to split the training data into a set used for optimization and a set used for validation.
During training, the performance of the network on the validation set is monitored and training is terminated when the performance on the validation set begins to drop.
Another method is dropout~\cite{srivastava_dropout:_2014}, where individual units of the network are randomly deleted during training.
The motivation for dropout is the idea that the network should be regularized by forming a weighted average of all possible parameter settings, with weights determined by their performance.
While this regularization is not feasible, removing units during training provides a reasonable approximation that performs well in practice.




\section{Theory}
\label{sec:theory}
The excellent performance of CNNs for various applications is undisputed, but the question of why remains mostly unanswered.
Here, we bring together a few different theoretical perspectives that begin to explain why CNNs are a good fit for solving inverse problems in imaging.

\subsubsection{Universal approximation}
We know that neural networks are universal approximators.
More specifically, a fully-connected neural network with one hidden layer can approximate any continuous function arbitrarily well provided that its hidden layer is large enough~\cite{hornik_approximation_1991}.
The result does not directly apply to CNNs because they are not fully connected, but, if we consider the network patch by patch, we see that each input patch is mapped to the corresponding output patch by a fully connected network.
Thus, CNNs are universal approximators for shift-invariant functions.
From this perspective, statements such as ``CNNs work well because they generalize X algorithm'' are vacuously true because CNNs generalize all shift-invariant algorithms.
On the other hand, the notion of universal approximation tells us what the network can learn, not what it does learn, and comparison to established algorithms can help guide our understanding of CNNs in practice.

\subsubsection{Unrolling}
The most concrete perspective on CNNs as generalizations of established algorithms comes from the idea of unrolling, which we discussed in Section~\ref{sec:arch}.
The idea originated in \cite{gregor_learning_2010}, where the authors unrolled the ISTA algorithm for sparse coding into a neural network.
This network is not a typical CNN because it includes recurrent connections, but it does share the alternating linear/nonlinear motif.
A more general perspective is that nearly all state-of-the-art iterative reconstruction algorithms alternate between linear steps and pointwise nonlinear steps,
so it follows that CNNs should be able to perform similarly well given appropriate training.
One refinement of this idea comes from \cite{jin2017deep}, which establishes conditions on the forward model, $H$, that ensure that the linear step of the iterative method is a convolution.
All of the inverse problems surveyed here meet these conditions, but the theory predicts that certain inverse problems, e.g. structured illumination microscopy, should not be amenable to reconstruction via CNNs.
Another refinement concerns the popular rectified linear unit (ReLU) employed as the non-linearity by most CNNs: results from spline theory can be adapted to show that combinations of ReLUs can approximate any continuous function.
This suggests that the combinations of ReLUs usually employed in CNNs are able to closely approximate the proximal operators used in traditional iterative methods.

\subsubsection{Invariance}
Another perspective comes from work on scattering transforms, which are cascades of linear operations (convolutions with wavelets) and nonlinearities (absolute value)~\cite{mallat_understanding_2016} with no combinations formed between the different channels.
This simplified model shows invariance to translation and, more importantly, to small deformations of the input (diffeomorphisms).
CNNs generalize the scattering transform, giving the potential for additional invariances, e.g., to rigid transformations, frequency shifts, etc.
Such invariances are attractive for image classification, but more work is needed to connect these results to inverse problems.

\section{Critiques}
\label{sec:critiques}
While the papers we have surveyed present many reasons to be optimistic about CNNs for inverse problems, we also want to mention a few general critiques of the approach.
We hope these can be useful points to think about when writing or reviewing manuscripts in the area, as well as jumping-off points for future research.

\subsubsection{Algorithm Descriptions and Reproducibility}
When planning this survey, we aimed to measure quantitative trends in the literature, e.g., to plot the number of training samples versus the number of parameters for each network.
We quickly discovered this is nearly impossible.
Very few manuscripts clearly noted the number of parameters they were training, and only some provided a clear-enough description of the network for us to calculate the value.
Many more included a figure of network architecture along the lines of Figure~\ref{fig:cnn_str}, but without a clear statement of the dimensions of each layer.
Similar problems exist in the description of the training and evaluation procedures, where it is not always clear whether the evaluation data comes from simulation or from a real dataset.
As the field matures, we hope papers converge on a standard way to describe network architecture, training, and evaluation.

The lack of clarity presents a barrier to the reproducibility of the work.
Another barrier is the fact that training often requires specialized or expensive hardware.
While GPUs have become more ubiquitous, the largest (and best-performing) CNNs remain difficult for small research groups to train.
For example, the CNN that won the ImageNet Large-Scale Visual Recognition Challenge in 2012 took ``between five and six days to train on two GTX 580 3GB GPU''~\cite{krizhevsky_imagenet_2012}.

\subsubsection{Robustness of Learning}
The success of any CNN-based algorithm hinges on finding a reasonable solution to the learning problem, \eqref{eq:learn_reg}.
As stated before, this is a non-convex problem, where the best solution we can hope for is to find one of many local minima of the cost.
This raises questions about the robustness of the learning to changes in the initialization of parameters and the specifics of the optimization method employed.
This is in contrast to the typical convex formulations of inverse problems, where the specifics of the initialization and optimization scheme provably do not affect the quality of the result.

The uncertainty about learning complicates the comparison of any two CNN-based methods.
Does A outperform B because of its superior architecture, or simply because the optimization of A fell into a superior local minimum?
As an example of the confusion this can cause, \cite{zhao_loss_2017} shows, in the context of denoising, super-resolution, and JPEG deblocking, that a network trained with the $l_1$ cost function can outperform a network trained with the $l_2$ cost function \emph{even with regard to the $l_2$ cost}.
In the authors' analysis of this highly disturbing result, they attribute it to the $l_2$ learning being stuck in a local optimum.
Regardless, the vast majority of work relies on the $l_2$ cost, which is computationally convenient and provides excellent results.

There is some indication that large networks trained with lots of data can overcome this problem.
In \cite{choromanska_loss_2015}, the authors show that larger networks have more local minima, but that most local minima are equivalent in terms of testing performance.
They also identify that the global minima likely correspond to parameter settings that overfit the training set.
More work on the stability of the learning process will be an important step towards wider acceptance of CNNs in the inverse problem community.

More generally, how sensitive are the results of a given experiment to small changes in the training set, network architecture, or optimization procedure?
Is it possible for the experimenter to overfit the testing set by iteratively tweaking the network architecture (or the experimental parameters) until state-of-the-art results are achieved?
To combat this, CNN-based approaches should provide carefully-constructed experiments with results reported on a large number of testing images.
Even better are competition datasets where the testing data is hidden until algorithm development is complete.

\subsubsection{Can We Trust the Results?}
Once trained, CNNs remain non-linear and highly complex.
Can we trust reconstructions generated by such systems?
One way to look at this is to evaluate the sensitivity of the network to noise: ideally, small changes to the input should cause only small changes to the output;
data augmentation during training can help achieve this.
Similarly, demonstrating generalization between datasets (where the network learns on one dataset, but is evaluated on another) can help improve confidence in the results by showing that the performance of the network is not dependent on some systematic bias of the dataset.

A related question is how to measure the quality of the results.
Even if a robust SNR improvement can be demonstrated, practitioners will inevitably want to know, e.g., whether the resulting images can be reliably used for diagnosis.
To this end, as much as possible, methods should be accessed with respect to the ultimate application of the reconstruction (diagnosis, quantification of biological phenomenon, etc.) rather than an intermediate measure such as SNR or SSIM.
While this critique can be made of any approach to inverse problems, it is especially relevant for CNNs because they are often treated as black boxes, and because the reconstructions they generate are plausible-looking by design, hiding areas where the algorithm is less sure of the result.

\section{Next Steps}
\label{sec:future}
We have so far given a small look into the wide variety of ways that researchers have applied CNNs to solve inverse problems in imaging.
Because CNNs are so powerful and flexible, we believe there remains plenty of room to create even better systems.
In this final section, we suggest a few directions that this future research might take.

\subsubsection{Biomedical Imaging}
CNNs have so far been applied most to inverse problems where the measurements take the form of an image and where the measurement model is simple, and less so for CT and MRI, which have relatively more complicated models.
A search on \url{arXiv.org} reveals dozens more CT and MRI papers submitted within the last few months, suggesting many incoming contributions in these areas.
We expect diffusion into other modalities such as PET, SPECT, optical tomography, TEM, SIM, ultrasound, super-resolution microscopy, etc. to follow.

Central to this work will be questions of how best to combine CNNs with knowledge of the underlying physics as well as direct and iterative inversion techniques.
Most of the surveyed works involve using a CNN to correct the artifacts created by a direct or iterative methods, where it remains an open question what is the best such prereconstruction method.
One creative approach is to build the inverse operator into the network architecture as in \cite{yang_deep_2016}, where the network can compute inverse Fourier transform.
Another would be to use the back projected measurements, $H^T y$, which at least take the form of an image and could reduce the burden on the CNN to learn the physics of the forward model.
CNNs could be deployed in a variety of other ways here, too, e.g. using a CNN to approximate a high quality, but extremely slow reconstruction method.
With enough computing power, a training set could be generated by running the slow method on real data, and, once trained, the resulting network could provide very fast and accurate reconstructions.

\subsubsection{Cross-Task Learning} 
In cross-task learning (also called \emph{transfer learning}, though this can have other meanings as well), an algorithm is trained with one dataset and deployed on a different, but related, task.
This is attractive in the inverse problem setting because it avoids the costly retraining of the network when imaging parameters change (different noise levels, image dimensions, etc.), which may occur often.
Or, we could imagine a network that transfers between completely different imaging modalities, especially when training data for the target modality is scarce;
e.g., a network could train on denoising natural images and then be used to reconstruct MRI images.
Recent work has made progress in the direction by learning a CNN-based proximal operator which can be used inside an iterative optimization method for any inverse problem~\cite{zhang_learning_2017}*.

\subsubsection{Multidimensional Signals}
Modern inverse problems in imaging increasing involve reconstruction of 3D or 3D+time images.
However, most CNN-based approaches to these problems involve 2D inputs and outputs.
This is likely because much of the work on deep neural networks in general has been in 2D, and because of practical considerations.
Specifically, learning strongly relies GPU computation, but current GPUs have maximally 24 GB of physical memory.
This limitation makes training a large network with 3D inputs and outputs infeasible.



One way to overcome this issue is model parallelism, in which a large model is partitioned onto separable computers. 
Another is data parallelism, where it is the data that is split.
When used together, large computational gains are achieved~\cite{dean2012large}. 
Such approaches will be key in tackling multidimensional imaging problems.



\subsubsection{Generative Adversarial Networks and Perceptual Loss}
CNN-based approaches to inverse problems also stand to benefit from new developments neural network research.
One such development is the generative adversarial network (GAN)~\cite{goodfellow_generative_2014}, which may offer a way to break current limits in supervised learning.
Basically, two networks are trained in competition, the generator tries to learn a mapping between training samples, while the discriminator attempts to distinguish between the output of the generator and real data.
Such a setup can, e.g., produce a generator capable of creating plausible natural images from noise. 
The GAN essentially revises the learning formulation \eqref{eq:learn_reg} by replacing the cost function $f$ with another neural network.
In contrast to a designed cost function, which will be suboptimal if the assumed noise model is incorrect, the discriminator network will act as a learned cost function that correctly models the probability density function of the real data from.
GANs have already begun to be used for inverse problems, e.g., for super-resolution in \cite{ledig_photo-realistic_2016}* and deblurring in \cite{schawinski_generative_2017}.

A related approach is perceptual loss, where a network is trained to compute a loss function that matches human perception.
The method has already been used for style transfer and super-resolution~\cite{johnson_perceptual_2016}.
Compared to the standard Euclidean loss, networks trained with perceptual loss give better looking results, but do not typically improve the SNR.
It remains to be seen whether these ideas can gain acceptance for applications such as medical imaging, where the results must be quantitatively accurate.


\bibliographystyle{IEEEtran}
\bibliography{biblio_final}


\begin{IEEEbiography}{Michael McCann} (S'10-M'15) received the B.S.E. in biomedical engineering in 2010 from the University of Michigan and the Ph.D. degree in biomedical engineering from Carnegie Mellon University in 2015.
He is currently a scientist with the Laboratoire d'imagerie biom\'{e}dicale and Centre d'imagerie biom\'{e}dicale,
\'{E}cole polytechnique f\'{e}d\'{e}rale de Lausanne (EPFL), where he works on X-ray CT reconstruction.
His research interest centers on developing signal processing tools to answer biomedical questions.
\end{IEEEbiography}%

\begin{IEEEbiography}
{Kyong Hwan Jin}
 received the B.S. and the integrated M.S. \& Ph.D. degrees from the Department of Bio and Brain Engineering, KAIST - Korea Advanced Institute of Science and Technology, Daejeon, South Korea, in 2008 and 2015, respectively. He was a post doctoral scholar in KAIST from 2015 to 2016.
He is currently a post doctoral scholar in the Biomedical Imaging Group, \'{E}cole polytechnique f\'{e}d\'{e}rale de Lausanne (EPFL), Switzerland. His research interests include low rank matrix completion, sparsity promoted signal recovery, sampling theory, biomedical imaging, and image processing in various applications.
\end{IEEEbiography}

\begin{IEEEbiography}
{Michael Unser} (M'89-SM'94-F'99) is professor and director of EPFL's Biomedical Imaging Group, Lausanne, Switzerland.
His primary area of investigation is biomedical image processing.
He is internationally recognized for his research contributions to sampling theory, wavelets, the use of splines for image processing, stochastic processes, and computational bioimaging.
He has published over 250 journal papers on those topics.
He is the author with P. Tafti of the book \emph{An introduction to sparse stochastic processes}, Cambridge University Press 2014. From 1985 to 1997, he was with the Biomedical Engineering and Instrumentation Program, National Institutes of Health, Bethesda USA, conducting research on bioimaging. Dr. Unser has held the position of associate Editor-in-Chief (2003-2005) for the IEEE Transactions on Medical Imaging. He is currently member of the editorial boards of SIAM J. Imaging Sciences, IEEE J. Selected Topics in Signal Processing, and Foundations and Trends in Signal Processing. He is the founding chair of the technical committee on Bio Imaging and Signal Processing (BISP) of the IEEE Signal Processing Society. Prof. Unser is a fellow of the IEEE (1999), an EURASIP fellow (2009), and a member of the Swiss Academy of Engineering Sciences. He is the recipient of several international prizes including three IEEE-SPS Best Paper Awards and two Technical Achievement Awards from the IEEE (2008 SPS and EMBS 2010).
\end{IEEEbiography}

\end{document}